# The regularized blind tip reconstruction algorithm as a scanning probe microscopy tip metrology method

**G Jóźwiak [1], A Henrykowski [1], A Masalska [1], T Gotszalk [1], I Ritz [2], H Steigmann[2]**
[1]Faculty of Microsystem Electronics and Photonics, Wroclaw University of Technology, ul. Janiszewskiego 11/17, 51-627 Wrocław
[2]Fraunhofer Institute for Non-Destructive Testing, Maria-Reiche-Strasse 2, D-01109 Dresden

E-mail: grzegorz.jozwiak@pwr.wroc.pl

**Abstract.**The problem of an accurate tip radius and shape characterization is very important for determination of surface mechanical and chemical properties on the basis of the scanning probe microscopy measurements. We think that the most favorable methods for this purpose are blind tip reconstruction methods, since they do not need any calibrated characterizers and might be performed on an ordinary SPM setup. As in many other inverse problems also in case of these methods the stability of the solution in presence of vibrational and electronic noise needs application of so called regularization techniques. In this paper the novel regularization technique (Regularized Blind Tip Reconstruction - RBTR) for blind tip reconstruction algorithm is presented. It improves the quality of the solution in presence of isotropic and anisotropic noise. The superiority of our approach is proved on the basis of computer simulations and analysis of images of the Budget Sensors TipCheck calibration standard. In case of characterization of real AFM probes as a reference method the high resolution scanning electron microscopy was chosen and we obtain good qualitative correspondence of both methods.
Keywords: scanning probe microscopy, blind tip reconstruction, regularization
PACS:

## 1. Introduction

In many fields of investigation which use SPM methods the interaction between an object and a tip is crucial and the tip shape and size strongly affects results. Also the method of estimation tip shape may make possible use SPM methods on new research areas. One of these are pull-off (adhesion) force measurements carried out by atomic force microscopy (AFM) techniques which have shown promise for determining solid surface free energy values at a submicrometer scale [1]. Such measurements might help with understanding behavior of industrially and medically important powders and its dependence on relative humidity [2] [3]. They play also an important role in studying of cancer cells giving the information about ligand receptor interaction [4]. Using very sharp tip to adhesion force measurements makes possible observation of hydrophobic interactions corresponding to only 25 chemical groups [5]. Such interactions mediate various crucial biological events, including protein folding, membrane fusion and cell adhesion. These forces are also important because they are believed to be one of the driving forces for the adhesion of pathogens to surfaces of tissues. The calculation of surface free energy from adhesion force measurements is quite straightforward and involves normalization of the measured adhesion forces by the contact area to field the work of adhesion. The models derived by Derjaguin, Muller and Toporov (DMT model) for very small, rigid probes interacting with smooth, rigid substrates of low surface energy [6] and Johnson– Kendall–Roberts (JKR model) for larger probes in systems with rather high surface energies [7], are the most frequently used in the analysis of pull-off forces. According to these two models, the measured adhesion force and the work of adhesion are related to each other by probe apex radius of curvature. Some authors [8] think that the AFM pull-off force measurements have not been widely accepted for measurements of solids because of an irreproducible nature of the data generated by this approach. Factors such as varying surface roughness and heterogeneity characteristics of both probes and substrates, and varying loads/deformation of the tip and substrate are among the major reasons for widely scattered data, even for the same set of materials used in a single experiment. A strict control of all of these conditions should be recognized to improve the precision of the results. In most of cited papers the AFM tip radius is an important factor not only in a problem of reconstruction of true surface geometry but also in case of modeling of complex biochemical interactions. The AFM probe shape is also important in measurements of surface mechanical properties for example Young modulus [9] and we think that the easy to handle and stable method of tip characterization provides a very useful tool improving the researches in this area.

The tip reconstruction methods might be divided into three groups: tip reconstruction by means of the tip characterizer with known geometry, blind tip reconstruction and tip measurements by means of the high resolution scanning electron microscopy (SEM). The first group suffers from characterizer imperfections. It enables characterization of tip with size for which characterizer uncertainty is negligible. On the other hand the SEM measurements are not convenient due to necessity of tip transporting from one device to another. Moreover the SEM systems with sufficient resolution are very expensive instruments and because of this they are not common in SPM laboratories.

We suppose that blind tip reconstruction methods have the best perspective, since they require only rough surface as characterizer and sophisticated numerical procedure. The numerical procedures for the blind tip reconstruction (BTR) were published in the same time by few authors [10,11,12] and later by Bakucz in [13]. We think that the most popular method, mainly due to very clear algorithm presentation [14] including program code in C language, was the method proposed by Villarubia in [10] (in this paper we will denote it as VBTR). This algorithm is also the part of two popular SPM image analysis software SPIP and Gwyddion.

The Villlarubia's algorithm works excellent on the simulated data but it is rarely used as a tool for real tip characterization. The main reason of such situation is its high sensitivity to noise. To ensure the stability of this procedure Villarubia proposed in [14] some regularization parameter changing the algorithm sensitivity to noise, of course at the cost of the resolution which is typical for many inverse problems. The procedure of determination of this parameter is not enough clear to be used by an average AFM operator as in case of people dealing with biochemical measurements. In this paper we investigate carefully this regularization technique and we propose the substantial improvements that might help to popularize this technique in a field of chemical force microscopy. Moreover we point out an important problem of noise directionality that significantly disturbs the regularization process.

## 2. Methodology and experimental setup

### 2.1. Basic definitions

In this section the necessary definitions and notations are introduced. We introduce only those definitions which are used in presented formulas. The thorough introduction would be too long and out of the scope of this paper. So for more details we refer the reader to the paper [10].

The utilized symbols coming from mathematical morphology and set theory enable convenient representation of useful theorems. Uppercase letters denotes a set of points in the $\mathbf{R}^3$ space. Lowercase bold letters denote a vector in the same space. Lowercase normal letters denote surfaces meant as two variable functions. An union of two sets $A$ and $B$ is expressed as $A \cup B$ while their intersection is denoted by $A \cap B$. A set translation by vector $\mathbf{d}$ is stood for $A + \mathbf{d}$. The presented methods use also so called dilation operation being defined as

$$A \oplus B = \bigcup_{b \in B} (A + \mathbf{b})$$ (1)

This operation is strictly connected with AFM imaging. In fact an AFM image is a measured surface dilated by an inverted tip.

### 2.2. Villarubia's method

Practitioners of SPM are well aware that image protrusions are broadened replicas of those on the specimen. However, it is only convention which determines which of the two objects being scanned across one another is the tip and which is the specimen. We are equally entitled to regard features on the image as broadened replicas (albeit inverted) of the tip. The VBTR method might be expressed in terms of mathematical morphology as

$$P_{i+1} = \bigcap_{x \in I} \left[ (I - \mathbf{x}) \oplus P_i'(\mathbf{x}) \right] \cap P_i, \text{ with } P_i'(\mathbf{x}) = \left\{ \mathbf{d} \mid \mathbf{d} \in P_i \text{ and } 0 \in I - \mathbf{x} + \mathbf{d} \right\}, \quad (2)$$

where $P_i$ is a tip reflected through an origin of a coordinate system at $i$-th iteration and $I$ is an image i.e. result of AFM measurements. This equation might be expressed by tip and image surfaces as follows:

$$p_{i+1}(\mathbf{x}) = \min_{\mathbf{x}' \in D_I} \left\{ \max_{\mathbf{d} \in D_{P'}} \left\{ \min^* \left[ i(\mathbf{x} + \mathbf{x}' - \mathbf{d}) + p_i(\mathbf{d}) - i(\mathbf{x}'), p_i(\mathbf{x}) \right] \right\} \right\}, \quad (3)$$

where $h(\mathbf{z})$ means a value of a surface function $h$ at the point $\mathbf{z} = [z_x, z_y]$, $D_I$ and $D_{P'}$ are domains of the image and the tip respectively, *min* and *max* mean minimal and maximal value in the given domain and *min\*(arg1,arg2)* means minimal value from arguments *arg1* and *arg2*. The VBTR procedure is iterative method and Villarubia proved [10] that each iteration of Eq. (2) produces a result smaller than or equal to the preceding one, but that each $P_i$ remains larger than the actual tip. This convergence limit is the best estimate of the tip possible to obtain by blind reconstruction.

This method would work perfectly if an imaging process were ideal. The sad truth is that SPM images (especially if we used chemically functionalized tips) have many measurement artifacts that have not been taken into account. As a result, an image besides the information about a specimen and a tip encloses also a noise. The main problem arises from the fact that mathematics describing electrical or vibrational noise differs significantly from those used for a tip reconstruction. Though an effect of dilation and convolution operators is frequently similar, the mathematical operations are completely different. In a blind tip reconstruction algorithm not only the noise makes the estimated tip uncertain but it also makes it biased. In practice the noise causes the estimated tip too sharp (fig. 1a).

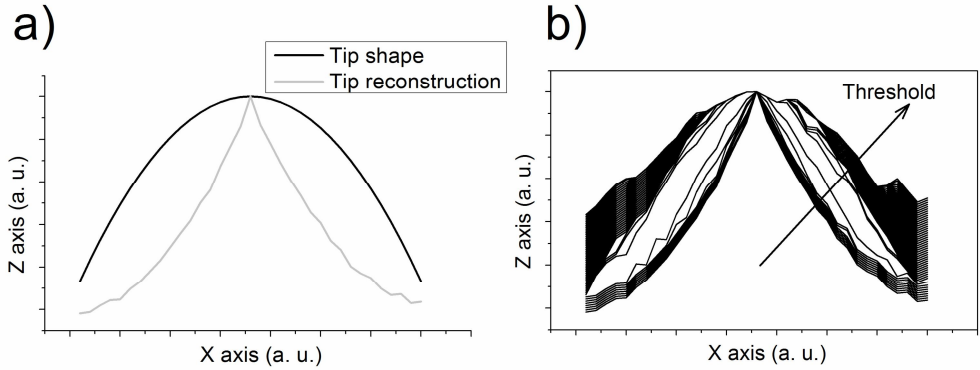

Figure 1. The role of noise in BTR estimator: a) the noise dependent bias, b) an influence of the regularization threshold (the arrow indicates the direction of threshold increase).

In order to overcome this difficulty in [14] the regularization parameter was introduced. This parameter establishes the level of inconsistency between the image and the tip which will be tolerated during the *min\** operation (3). In this case the equations (3) should be presented as follows:

$$p_{i+1}(x) = \min_{\mathbf{x}' \in D_I} \left\{ \max_{\mathbf{d} \in D_{P'}} \left\{ \min^* \left[ i(\mathbf{x} + \mathbf{x}' - \mathbf{d}) + p_i(\mathbf{d}) - i(\mathbf{x}') + t_H, p_i(\mathbf{x}) \right] \right\} \right\}, \quad (4)$$

where $t_H$ is the threshold value. Fig. 1b shows how the threshold affects a process of tip − surface intersection calculation. We have to realize that noise makes the analyzed image uncertain. Each image point might lay upper or lower with some probability. Since in a dilation operation we use *min* and *max* operations we need to assume that a true image lays above all points of a noisy image. This is the cause of an introduction of some threshold parameter that establishes how much the assumed image is above the noisy one. The optimal case is when the threshold is equal 5 times standard deviation of noise. The fig. 2a shows how an additive electrical noise might be treated by mathematical morphology. The VBTR method uses this threshold parameter to add a positive feedback compensating bias caused by the noise. The main problem with this regularization scheme arises from the fact that usually we do not know the noise standard deviation. Of course we can try to estimate it from an image but such estimation depends strongly on a shape of an investigated surface. Dongmo proposes a method [15] that utilizes tip shape volume for threshold determination. It is a fact that when threshold becomes sufficiently greater than noise a tip shape becomes much blunter and their volume suddenly increases. Dongmo proposes doing reconstruction starting from low to high threshold values with some step. The threshold value above a sharp transition, where the changes in a tip shape with increasing threshold become small again, is regarded as optimal. Such procedure is not very precise, because the threshold value significantly depends on the step length and what is even more important if the step length becomes lower we obtain worse results instead of better ones. We will show by an experiment that this regularization mechanism might give threshold values that are far away from optimum.

*2.3 Proposed algorithm*

*2.3.1. Estimation procedure.* In this subsection we present our proposition of a regularization mechanism which in our opinion is much more consistent in treating image points as uncertain. Moreover, we are going to prove that our solution is much more stable as compared with VBTR. Villarubia introduced threshold parameter as a mechanism for compensation of the negative bias caused by noise. We propose to look at this problem from a different point of view. If we treat each image point as uncertain we conclude that for correctly determined threshold value the image without noise is enclosed between two parallel copies of the noisy image which are the threshold distance apart (fig. 2a).

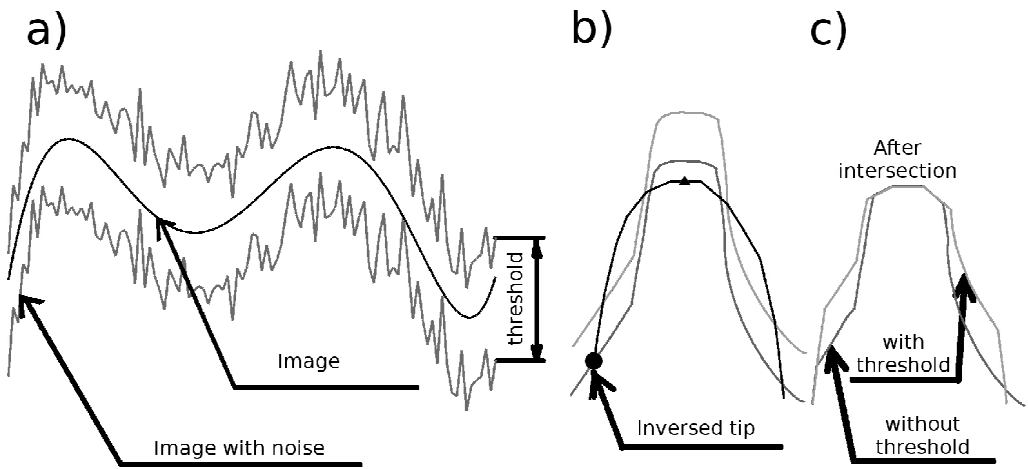

Figure 2. A VBTR regularization mechanism: a) an illustration of relations between the threshold, noisy image and image without noise, b) an illustration of the case when a tip touches a surface at the point marked by the circle, c) a tip-image intersection with and without the threshold.

This point of view reveals that we have to take into account the fact, that the set $P_i'(\mathbf{x})$ is also established on the basis of a noisy image. The VBTR procedure ignores this fact which is presented in the figure 3 , where the tip shape presented in the figure 3b is not included in *max* operation in equation (3), because the tip apex (indicated by triangle) is above the image (fig. 3a). This might be corrected if the formula on $P_i'(\mathbf{x})$ were expressed as follows:

$$P_i'(\mathbf{x}) = \left\{ \mathbf{d} \mid \mathbf{d} \in P_i \; and \; 0 \in I - \mathbf{x} + \mathbf{d} + t_H \right\}. \tag{5}$$

In the experimental part of this paper we show that such determination of $P_i'$ set improves substantially regularization process.

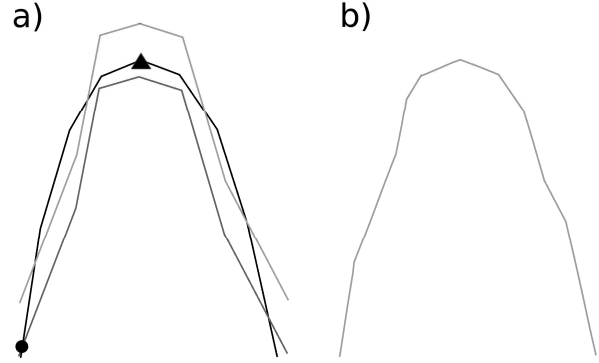

Figure 3. Checking if a tip apex is under an image: a) the black line is a tip, the gray is an image and the light gray line shows the image increased by the threshold (VBTR rejects this case because the tip apex is over the image, regardless the image is uncertain), b) the tip shape (it is an intersection between the image risen by the threshold and the an actual tip shape) not included to *max* operation in the equation (3).

*2.3.2. Noise anisotropy.* In our solution we also take into consideration an effect of noise directionality that significantly disturb a regularization process. It makes an estimated tip apex flat in one of the directions. It would be even acceptable if this flattening occurred in the direction of the noise. Unfortunately it appears in the direction that might be reconstructed with better resolution. The main reason of this problem is that we determine one threshold value on the basis of a tip volume. In order to solve this problem, we propose to determine the threshold values for each direction separately. It might be done on the basis of an area of the tip cross sections along $X$ (fast scan axis) and $Y$ (slow scan axis) direction. In this way we obtain two reconstructions, each one obtained for a different threshold value. The tip obtained for the lower threshold value has correct shape along one direction (usually fast $X$ direction) and is too sharp along the second one (usually slow $Y$ direction)(fig. 4a). The tip obtained for the higher threshold value has correct shape along the slow scan axis but is significantly broadened along the fast scan axis (fig. 4b). If we consider the tip as a matrix, we might conclude that for the first tip each row has a proper shape but its position along the $Z$ axis is disturbed by the directional noise. We can correct the row positions on the basis of the cross section of the tip obtained for the higher threshold value (fig. 4c). In such a way we recover the maximum information from the image and simultaneously, ensure proper regularization. If the tip $p_X$ is obtained for the threshold $t_{HX}$, determined on the basis of a cross section along fast direction and the tip $p_Y$ is obtained for the threshold $t_{HY}$, determined on the basis of a cross section along slow direction and additionally $c_x$ and $c_y$ are coordinates of the tip apex, the process of tips combination is expressed by following equation:

$$p(x,y) = p_X(x,y) + p_Y(c_x,y) - p_X(c_x,y),$$

(6)

where $x$ and $y$ are coordinates of tip points.

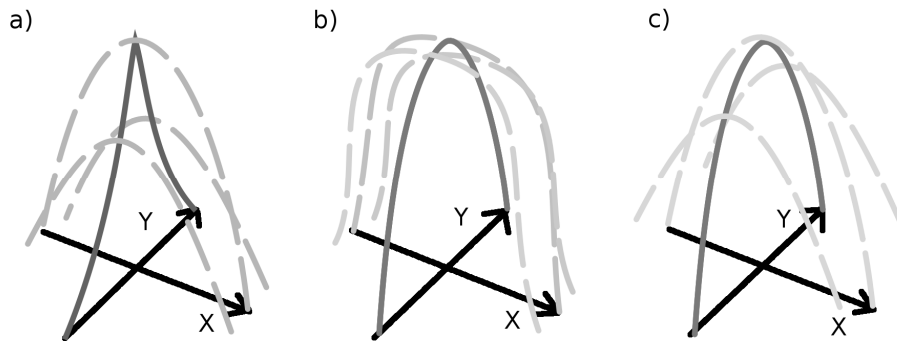

Figure 4. The method of tip combining: a) the BTR tip obtained for threshold $t_{HX}$, b) the BTR tip obtained for threshold $t_{HY}$, c) the combined tip (rows of the tip $p_X$ which are translated according to cross section of the tip $p_Y$).

*2.3.3. Initial tip shape.* The presence of additive noise in AFM images increases values of threshold parameters. It means that some part of an initial tip apex is untouched by reconstruction procedure. If we start with a flat initial tip a part of this flat region remains unchanged and the estimated tip will have flat apex. But from the theory we have better estimation of this part of tip. We can use any spiky element from an image and use it as an initial tip. If the blind tip reconstruction algorithm does not change some part of the initial tip due to a high threshold value we still have a rough approximation of this tip region. Because of that, we have developed the procedure that builds the initial tip shape on the basis of most spiky element of an image. This element is chosen on the basis of a specially designed criterion function. The candidates for the initial tip must have the maximum at the tip center. For each part of the image meeting this criterion we calculate the measure of its sharpness as follows

$$M = \sum_{i,j} \exp\left[-a\sqrt{(i-i_s)^2 + (j-j_s)^2}\right]\left[I(i_s, j_s) - I(i, j)\right], \qquad (6)$$

where $i_s$, $j_s$ are indices of the center of the tip, $I(i, j)$ is the image height at the point $(i,j)$ and $a$ is defined as $-0.125log_2W$, where $W$ is the assumed tip width. The maximum of four image parts with the greatest $M$ value is chosen as an initial tip shape.

The initial tip shape determined in such a way is disturbed by noise and a noise influence is neither corrected nor worsen in regularization process. So this interference is random and does not bias the solution but only decrease precision of an estimated tip. In tip radius estimation procedure for example, when a tip is fitted by a parabola, such random disturbances become a fitting error and practically do not affect the radius of curvature of the tip.

*2.3.4. The Slope detection procedure.* In VBTR method a slope of volume vs. threshold relation is detected on the basis of a numerical derivative. We suppose that using a numerical derivative makes the solution sensitive to the step length. Moreover in VBTR the optimum threshold value is determined on the basis of the rule of thumb saying that we should choose the threshold for which changes of the tip volume become again small. This statement is not precise enough i.e. in fig 6c we see small changes starting from triangle point up to the last point (this part of the curve is concave so the numerical derivative decrease monotonically) while the optimum point denoted by the circle lays in the middle of this interval. Instead of using numerical derivative the RBTR slope detection procedure fits the volume vs. threshold function by 3 straight, joined lines. The fitting procedure is iterative. We start from three lines for which second line join the extreme points connecting the maximum volume slope. The first and the third have to form continues function with the second one and minimize the norm

$$L_2 = \int_0^{t_{Hmax}} (L(t_H) - V(t_H))^2 dt_H, \qquad (7)$$

,where $L(t_H)$ is the 3 line function and $V(t_H)$ is the linear interpolation of the volume vs. threshold relation. Next we iteratively move the points connecting the lines and investigate the $L_2$ value. If the $L_2$ does not decrease, we stop the procedure. The optimal threshold value is the nearest to the point connecting the second and the third line.

## 3. Experiment description and discussion

### 3.1. Reconstruction on the basis of the simulated data

The images used in this part of experiment were obtained by simulation of scanning process of spiky surface that consists of sharp cones with random amplitude and tilt. We assumed that surface is scanned with elliptic paraboloid tip. This image is perturbed by two types of noise. The first type is simple Gaussian noise. Figure 5a illustrates the analyzed image and figure 5b shows the tip utilized for simulation of the scanning process. The presented images are visualized by TOPOGRAF, the homemade image analysis software. The SNR counted as ratio of image $S_q$ roughness parameter to noise standard deviation is about 10. Such image was analyzed by VBTR and RBTR algorithms. As a results we have obtained tips presented in figures 5c (VBTR) and 5d (RBTR). It is shown that RBTR algorithm has estimated tip shape much better than VBTR.

To examine this case more thoroughly we present the cross-section of tip shapes obtained by both methods with relation to regularization parameter values. The RMS value is defined as follows:

$$RMS = \sqrt{\frac{\sum_{i=1}^{N_R}\sum_{j=1}^{N_C}\left(P_S(i,j) - P_R(i,j)\right)^2}{N_R N_C}} \; , \tag{8}$$

where $P_S$ is a simulated tip and $P_R$ reconstructed, $N_R$, $N_C$ are number of rows and columns respectively. We can see that in case of both methods the cross section area increase suddenly but in case of VBTR method this jump appears too early. This point is denoted by triangle in the figure 6c. For this point the RMS standard error between the original tip and the estimated one is about 1.306. Comparing to the standard error of RBTR tip, equal to 0.306, it is more than four times greater. Since the RBTR uses much more sophisticated slope detection algorithm we check what happens if we use the VBTR with the RBTR slope detection procedure. This case is marked by diamond in the figure 6c. The application of our slope detection procedure improves the result but it is still three times worse than RBTR. The optimal threshold for which the results of the RBTR and the VBTR are comparable is pointed by the circle in the figure 6c and we can see that this point is ordinary and cannot be found on the basis of a cross-section vs. threshold relation. That fact means that problem is inside the reconstruction procedure and not the slope detection one.

The cross section vs. threshold relation in the RBTR is much more reliable. In this case the slope is steep and the procedure based on three lines fitting gives a correct threshold value. Moreover the slope detection algorithm of the RBTR method, in contrast to simple VBTR rule, is almost independent on the step length of the regularization threshold and we need not to consider if the changes of tip volume are sufficiently small or not.

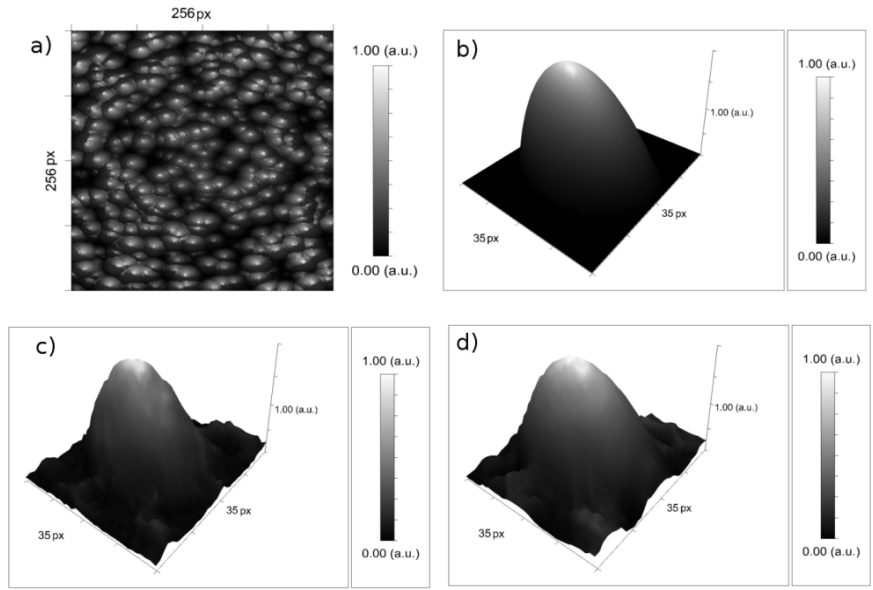

Figure 5 The results of a blind tip reconstruction on the basis of a simulated image: a) simulated noisy image, b) tip used for imaging process, c) tip estimated by RBTR method, d) tip estimated by VBTR method.

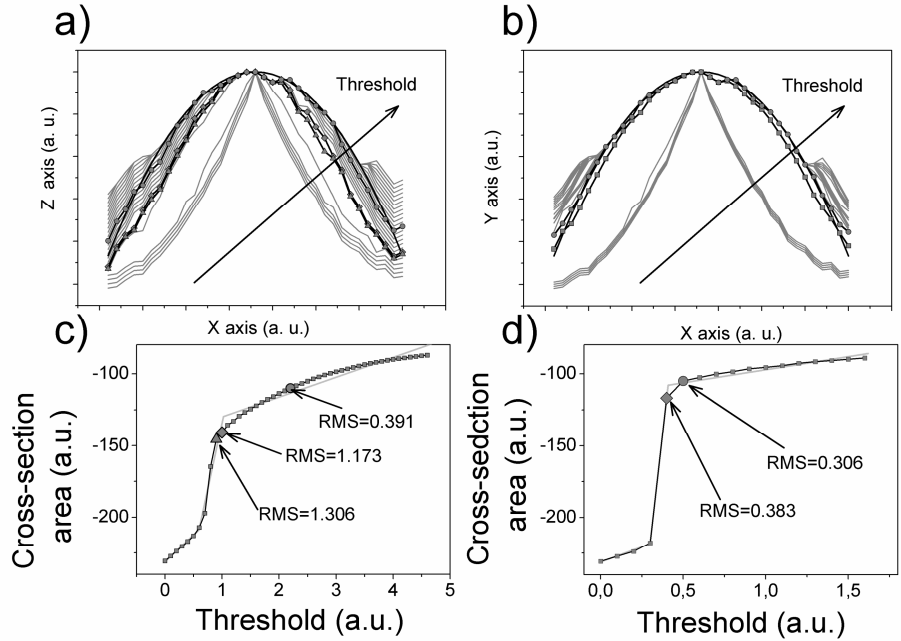

Figure 6. Illustration of regularization mechanism in case of isotropic Gaussian noise: a) cross-sections of tip shapes obtained with VBTR method for different threshold values (triangle cross section is obtained for threshold determined by VBTR method, diamond cross section is determined on the basis of VBTR method with RBTR slope detection procedure and circle cross section is determined for threshold value that gives minimum RMS value), b) cross-sections of RBTR tip shapes for different threshold values (diamond cross section is for the threshold below the value computed by RBTR slope detection procedure, circle cross section is for the optimal threshold value) , c) relation of VBTR cross-section area vs. threshold value, d) relation of RBTR cross-section area vs. threshold value.

The previous simulations use an isotropic Gaussian noise but in case of AFM measurements an anisotropic noise is often observed. It might be recognized as random horizontal lines interfering the AFM image (figure 7a). This noise significantly affects the regularization process of blind tip reconstruction algorithms. The noise visible in the figure 7a was simulated by adding horizontal lines having zero slope and a randomly determined intersection. The signal to noise ratio SNR, computed in the same manner as in the case of Gaussian noise, is equal to 24. The figure 7b shows cross sections along fast scan axis. The solid gray line shows a shape of the simulated tip. The solid black line shows the cross-section of the tip shape estimated on the basis of the cross-section area vs. threshold relation. The dotted line presents the cross-section of the tip reconstructed on the basis of the volume vs. threshold relation.

A tip cross-section in Y direction is not affected by noise anisotropy and from this reason threshold values achieved by the Y cross-section area vs. threshold relation and the volume vs. threshold relation are the same (the black solid line in the figure 7c). Therefore using a cross-section area vs. threshold relation improves a reconstructed cross-section in a direction orthogonal to direction of noise and this is what we want to achieve. If there is no noise in the given direction we want to reconstruct it perfectly.

In the paper [16], Todd and Eppell proposed a method dealing with noise directionality but they introduce a threshold function of two variables (thresholds) corresponding to noise levels along fast and slow scan direction. Similarly to classic VBTR approach, the values of these two thresholds are determined on the basis of the behavior of the tip volume. However in this case we have to know the volume of the tip for every point belonging to 2D grid of thresholds. It means that for each grid point we have to do blind tip reconstruction. The number of reconstructions depends on grid density. In our simulations we do reconstruction for 20 equidistant thresholds values. If we want to preserve this density in case of Todd and Eppell procedure we need to do 400 reconstructions. Since the blind tip reconstruction is the most time consuming part of the procedure, our solution is about 20 times faster than the procedure of Todd and Eppell.

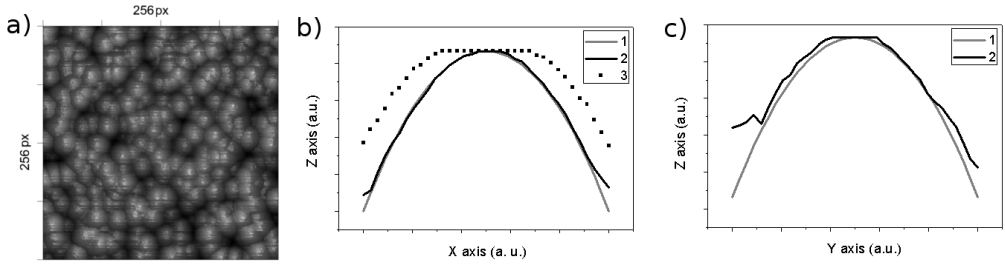

Figure 7. The analysis of simulated image and tip with an anisotropic noise: a) simulated noisy image, b) tip cross-sections along X (fast scan) axis (the first line is original tip shape, the second line shows cross section of tip estimated on the basis of cross section area vs. threshold relation, the third line presents cross section of tip estimated on the basis of volume vs. threshold relation), c) tip cross-sections along Y (slow scan) axis ( the first line shows original tip shape, the second line presents the cross section of the tip estimated on the basis of volume as well as of cross section area vs. threshold relations).

### 3.2. Reconstruction on the basis of AFM images of calibration standard

In this part of experiment the reconstruction process was done on the basis of AFM images of the Budget Sensors TipCheck sample. For AFM measurements we utilized Veeco Multimode AFM system with NanoScope V controller. The size of scanned area was 500x500 nm, sampled uniformly at the grid of 512x512 points. The sample was scanned by two Nanoworld tips from which one was new tip with tip radius about 10 nm as producer ensured. The second probe was worn tip with unknown tip radius. Figure 8 shows the corresponding AFM images. The images are interfered and the noise is apparently anisotropic. Therefore we use RBTR method designed for anisotropic noise.

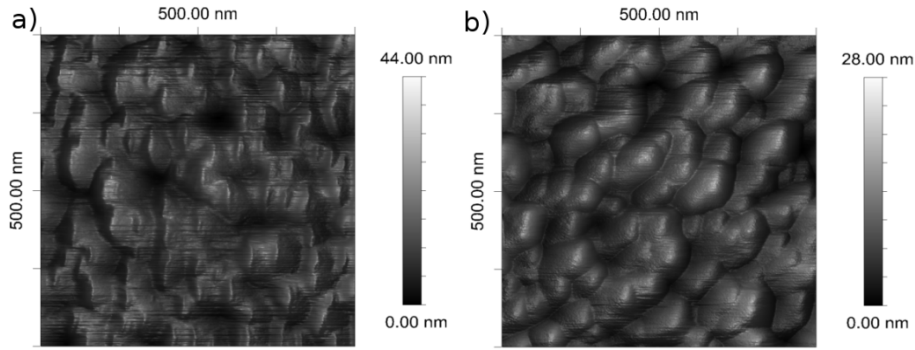

Figure 8. The AFM images of Budget Sensors TipCheck sample: a) scanned by new tip, b) scanned by worn tip.

As a reference method we use scanning electron microscopy (SEM). The measurement were performed with Carl Zeiss Auriga system. Figure 9 shows the results obtained by both (SEM and RBTR) methods for two investigated tips. To compare the results we have superimposed the SEM image and RBTR shape cross-section in X direction, which should correspond approximately. To compare the results quantitatively the corresponding tip radiuses have been calculated. Surprisingly, we obtain different results for blunt and sharp tip. The results are as follows: $R_{RBTR}$=8 nm and $R_{SEM}$=14 nm for the sharp tip while for the blunt $R_{RBTR}$=55 nm and $R_{SEM}$=33 nm. So for the sharp tip RBTR method leads to sharper tip while for the blunt to blunter if we compare them to SEM. The results achieved for worn tip are more reasonable because AFM images are noisy and give only upper boundary of the tip, so it might be sharper. To improve the correspondence of the results we need to decrease the noise in the image by application of one of filtration techniques. On the other hand, results obtained for sharp tip suggest that RBTR method gives more information about tip shape than SEM. In order to proof the correctness of RBTR procedure we searched the AFM image for structures with radius of curvature equal or lower to 8 nm. We found at least 7 structures in the image of the TipCheck sample that have the radius of curvature about 8 nm or even less and do not look as artifacts. This suggests that the RBTR procedure works correctly. A potential source of discrepancy between SEM and RBTR results might be the fact that SEM images gives only projections of the tip. Since we do not know the angle between the probe and the surface, we actually do not know how to correctly position the detector of the SEM system to compare both results. Moreover, in order to obtain tip radius we fitted the tip shapes by parabola. The size of the tip part that is fitted by parabola is different for the RBTR tip and the SEM tip, because smaller part of RBTR tip corresponds well with the parabola. If we take the same part of RBTR tip as for SEM tip, the RBTR tip radius will be about 12 nm but the parabola will not fit very well to the RBTR tip. Concluding, to obtain quantitative agreement of the SEM and RBTR tips further investigations are necessary.

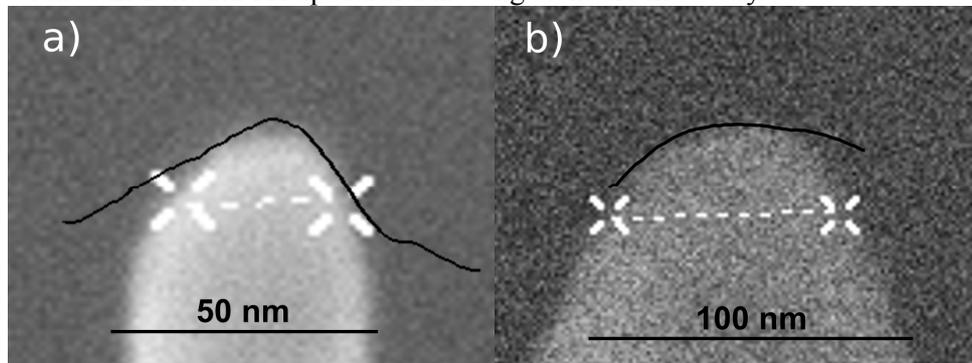

Figure 9. SEM images and RBTR cross sections along X axis of the: a) new and b) worn tip.

## 4. Conclusions

The conditions needed to be met by tip shape estimation procedure are presented in the paper. We think that such conditions might be met by blind tip reconstruction methods if the proper regularization mechanism is applied. We point out the problems with standard regularization technique based on investigation of tip volume vs. threshold relation. The proposed RBTR method improves the regularization by consequent treating an AFM image as uncertain. We also considered the case where the noise is anisotropic and propose solution that is much more effective with comparison to those proposed by Todd and Eppell in [16]. The RBTR method was tested on simulated images as well as on AFM images of the TipCheck sample. In case of the simulated data the proposed method shows that is superior to previously proposed regularization techniques. For AFM images the results obtained by RBTR and SEM methods are comparable but some inconsistencies needs further investigations.